\documentclass[a4paper,twocolumn,10pt]{article}
\usepackage{eurosis}
\usepackage{url}

\usepackage{cite}
\usepackage{amsmath,amssymb,amsfonts}
\usepackage{algorithmic}
\usepackage{graphicx}
\usepackage{textcomp}
\usepackage{xcolor}
\usepackage{enumitem}

\usepackage{pgfplots}
\pgfplotsset{compat=newest}
\usepackage{tikz}
\usepackage{xcolor}
\usepackage{multirow}
\usepackage[center]{caption}
\usepackage[integrals]{wasysym}

\newlength\figureheight 
\newlength\figurewidth 
\begin{document}

\title{ANALYZING THE COUPLING PROCESS OF DISTRIBUTED MIXED REAL-VIRTUAL PROTOTYPES}

\author{Peter Baumann & Lars Mikelsons & Dieter Schramm \\
	Oliver Kotte & Chair for Mechatronics & Chair for Mechatronics\\
Robert Bosch GmbH & University of Augsburg & University of Duisburg-Essen\\
\texttt{peter.baumann5@de.bosch.com} & Augsburg, Germany  & Duisburg, Germany \\
\texttt{oliver.kotte@de.bosch.com} & \texttt{lars.mikelsons@informatik.uni-augsburg.de} & \texttt{schramm@imech.de}}

\date{}

\maketitle

\thispagestyle{empty}


\begin{abstract}
The ongoing connection and automation of vehicles leads to a closer interaction of the individual vehicle components, which demands for consideration throughout the entire development process. 
In the design phase, this is achieved through co-simulation of component models. 
However, complex co-simulation environments are rarely (re-)used in the verification and validation phases, in which mixed real-virtual prototypes (e.g. Hardware-in-the-Loop) are already available.  
One reason for this are coupling errors such as time-delays, which inevitably occur in co-simulation of virtual and real-time systems, and which influence system behavior in an unknown and generally detrimental way.
This contribution introduces a novel, adaptive method to compensate for constant time-delays in potentially highly nonlinear, spatially distributed mixed real-virtual prototypes, using small feedforward neural networks.
Their optimal initialization with respect to defined frequency domain features results from a-priori frequency domain analysis of the entire coupled system, including coupling faults and compensation methods.
A linear and a nonlinear example demonstrate the method and emphasize its suitability for nonlinear systems due to online training and adaptation. As the compensation method requires knowledge only of the bandwidths, the proposed method is applicable to distributed mixed real-virtual prototypes in general.
\end{abstract}

\def\svgwidth{\columnwidth}
\section{INTRODUCTION}

Current trends in the automotive sector like connected and autonomous driving functions are leading to a closer coupling of different vehicle domains.
This is for example the case in the development of an emergency brake assistants through the interaction of longitudinal control and brake management.
In order to enable a time- and cost-efficient development of such cross-domain vehicle functions, the interactions of the domains must be considered at an early stage of the development process.
Simulation experts interconnect the different simulation models, by coupling various tools using the methods of co-simulation~\cite{Gomes.2018}.
This way complex cross-domain model-in-loop (MiL) co-simulations are implemented.
Since the modeling and integration effort to set up such co-simulations is high, the demand is coming up to use the same co-simulation environment not only during the design phase, but also in the verification and validation phase of the development process. 
In those phases, first components of the vehicle are available as real hardware on test benches.
Coupled to the existing MiL co-simulation, detailed mixed real-virtual prototypes are realized to test the hardware or software in open context under realistic (e.g. traffic) conditions.
But, due to the real-time requirement, couplings between simulation models and hardware (HiL) always come with coupling faults (e.g. time-delay or measurement noise).
If, in addition, the unchanged models from the MiL co-simulation are to be used, the coupling faults even increase, since the models can generally not be compiled on a real-time operating system, but run on a standard Windows PC. 
Furthermore, there are use cases for cross-company collaboration using mixed real-virtual prototypes, since the complexity of the systems is increasing and their handling requires a wide range of different competencies, which most companies do not have in house.
The additional distance between the coupled systems in those use cases further increases the coupling faults and their negative effect on the overall coupled system.

This contribution addresses the time-delays in distributed mixed real-virtual prototypes, by proposing a novel, adaptive compensation method combined with a detailed analysis of the dynamic effects a simulator distribution has on the overall system.
After covering related work from the fields of distributed mixed real-virtual prototypes and coupling fault compensation, the compensation method based on a feedforward neural network is presented.
The results of the following analysis of the overall system in frequency domain are then used for an optimal compensation method design. Finally, the applicability of the methodology on a nonlinear example is shown.

\section{RELATED WORK}
\label{related}
Recently the development of mixed real-virtual prototypes in the automotive sector gained some momentum through the release of the Distributed Co-Simulation Protocol (DCP)~\cite{Krammer2018}.
The DCP is designed to standardize the coupling of real-time or non real-time simulators and thus reducing the integration effort of spatially distributed prototypes.

In literature many examples for the transition from virtual to real testing using mixed real-virtual prototypes can be found. 
A toolchain for a seamless transition from a MiL co-simulation to heterogeneous HiL testing is presented \cite{Klein.2017}. Using this toolchain, an engine test bench is coupled to real-time vehicle dynamics and environment models.
Mixed real-virtual prototypes can also be used to incorporate the driver in the testing process, e.g. by investigating the interaction between humans and an automatic transmission in a driving simulator~\cite{Maas.2014}. 
Since test benches usually do not stand side by side, it is reasonable to use the internet when coupling them. 
Examples for mixed real-virtual prototype coupled via the internet can be found in~\cite{Ersal.2009} and~\cite{Schreiber.2018}. In~\cite{Baumann.2019} additionally the DCP is used for the integration.

In all mentioned examples, the focus is on implementing the coupling itself.
Coupling faults as delays and dropouts, specified in~\cite{Schreiber.2018}, are known but the use of compensation methods is not yet widespread.
In the scientific fields of telerobotics~\cite{Lawrence.1993} and networked control systems~\cite{Baillieul.2007} many methods are developed to compensate for these coupling faults, but the compensation is implemented in the controls themselves.
For mixed real-virtual prototypes, however, the physical models should not be modified, which is why the compensation must be implemented in the coupling signals~\cite{Stettinger.2017}. 
In~\cite{Stettinger.2014} a model-based-coupling method is presented which is meant for the usage for mixed real-virtual prototypes and is tested on an engine test bench. The parameters of two second order linear systems are identified online to compensate for the delay. 
A linear fourth order FIR filter is introduced in~\cite{Stettinger.2017}. 
Together with a recursive least squares algorithm as identification method, the filter is designed to cope with communication time delays, data-losses and noisy measurements.

However, there is no compensation method capable of representing nonlinear signal behavior. In addition, the influence of the compensation on the overall system should be predictable and verifiable and its parameterization should ideally be based on the dynamics of the coupled system itself.
\section{COMPENSATION METHOD}
\label{method}

This paper considers the coupling process between two distributively coupled systems (e.g. vehicle components) ``A'' and ``B'', at least one of which is a real-time system.
The communication or macro step size $\Delta T$ is the constant rate at which data is exchanged between the two systems.
The coupling faults that occur in a distributed coupling of mixed real-virtual prototypes are attributable to effects like communication time-delay, jitter, determinism and message loss~\cite{Schreiber.2018}. 
Since the time-delay is usually the most dominant fault in a distributed system, the others are neglected in this paper.
This simplifies the faults to a constant time-delay $\tau$ which represents the time between sending a message from ``A'' and receiving it at ``B'' and vice versa. The macro step size fixes its resolution, it holds $\tau = k \cdot \Delta T$ with $k \in \mathbb{N}$. 

\begin{figure}[h]
	\def\svgwidth{\columnwidth}
	\centerline{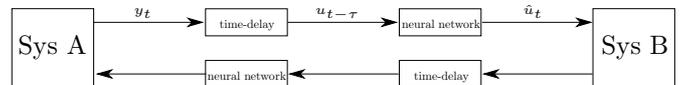}
	\caption{Overview of the Considered Coupled System}
	\label{fig:CouplingOverview_simple}
\end{figure}

To compensate for the time-delay, an algorithm is needed which is placed at each input $u$ of each system participating in the distributed real-time co-simulation. 
The compensation method extrapolates the delayed input $u_{t-\tau}$ to get a predicted input value $\hat{u}_{t}$ at time $t$.
In order to make the method applicable for as many simulation tools and real-time systems as possible, the extrapolation is signal-based and only the current as well as past values (and no derivatives) of the inputs are used for the extrapolation. Figure~\ref{fig:CouplingOverview_simple} gives an overview of the coupled system, the subscript of $y$, $u$ and $\hat{u}$ stand for the point in time at which the signal is evaluated.

For extrapolation, small feedforward neural networks are used here.
If the activation function of all network nodes is linear and a vector $\vec{u}$ consisting of $p$ past signal values is the input to the network, the output of the network reads as
\begin{equation}
\label{ANN_func}
\hat{u}_{t} = \vec{a}^T\vec{u} + b
\text{ with }  
\vec{a} = 
\begin{bmatrix}
a_{1}\\
a_{2} \\
\vdots \\
a_{p} 
\end{bmatrix} 
\text{and }
\vec{u} =     
\begin{bmatrix}
u_{t-\tau}\\
u_{t-\tau-\Delta T} \\
\vdots \\
u_{t-\tau - (p-1) \Delta T} 
\end{bmatrix} 
\text{.}
\end{equation}
The vector $\vec{a}$ and the scalar $b$ are calculated from the weights of the network.
As already stated by~\cite{Dorffner.1996}, this equals a linear autoregressive model, which is a generalization of commonly used extrapolation methods in co-simulation as zero-order-hold (constant extrapolation, ZOH) and first-order-hold (linear extrapolation, FOH). 
The recursive FIR-Filter for time-delay compensation in~\cite{Stettinger.2017} is based on the same function. 

The main advantage of using a feed forward neural network for the extrapolation is the fact that it can be extended to represent nonlinear signal behavior, by choosing a nonlinear activation function in the hidden layer(s).
If Rectified Linear Units (ReLU) or, to prevent parts of the network from ``dying'', leaky ReLUs are chosen as activation function, the network is enabled to switch between different configurations of the parameters $\vec{a}$ and $b$.
For example, a hidden layer with $n$ neurons with (leaky) ReLU activation implements $n^2$ different parameter sets of $\vec{a}$ and $b$ depending on which hidden node is active and which is not.
But between the switching points the network retains its linear behavior from equation \eqref{ANN_func}, which allows a detailed analysis of the compensation behavior, even in the nonlinear case.
Furthermore, the structure of the neural network allows using efficient algorithms for the initial training as well as an online adaption of the weights of the network.

\section{ANALYSIS OF THE COUPLING PROCESS}
\label{analysis}
The coupling faults influence the behavior of a distributed mixed real-virtual prototype significantly. 
It is not unlikely that an originally stable system gets unstable due to the coupling and the associated coupling faults.  
In order to increase the confidence in mixed real-virtual prototypes by making statements on robustness regarding the coupling faults, this section aims to analyze the entire closed-loop system including coupling faults and compensation method in frequency domain.
First, the transfer function of the coupling process including coupling faults and compensation method is calculated and second it is shown using an example how this transfer function can be utilized to analyze the influence of the coupling faults and of the compensation method on the closed-loop system.
Furthermore, it is shown in the next section that the compensation method can be initialized optimally based on this analysis.

\begin{figure}[h]
	\def\svgwidth{\columnwidth}
	\centerline{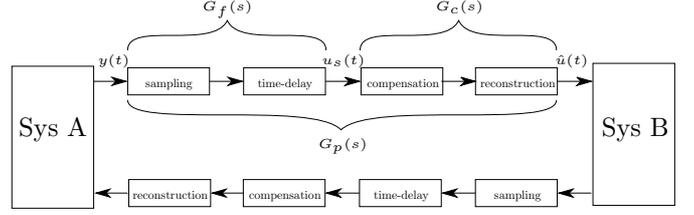}
	\caption{Components of the Real-Virtual Prototype}
	\label{fig:CouplingOverview}
\end{figure}

As already mentioned in~\cite{benedikt2019relaxing}, a detailed analysis of a co-simulation is also possible in time domain via a multi-rate approach, but due to the variable step size solvers usually used for physical models, the analysis would become very complex.
Instead, in this paper, the analysis is carried out in frequency domain which leads to the following assumptions:
\begin{enumerate}[noitemsep]
	\item  All sub-systems, their inputs and their outputs are assumed to be ideally time continuous. 
	It follows that the numerical errors made by a solver due to evaluating the system equations at discrete points in time are neglected. 
	The measurement error that occurs when reading out values of real-time systems (e.g. test benches) is also neglected. 
	\item The macro step size $\Delta T$ is constant and small enough to avoid aliasing in all coupling signals. 
	This is verified via the Nyquist-Shannon theorem with the condition $\bar{\omega} \Delta T << \pi$ for the ratio of the maximal bandwidth $\bar{\omega}$ of the coupling signal and $\Delta T$~\cite{Benedikt.2013}.
	\item The coupling faults are simplified to a constant time-delay $\tau$ (see previous section).  
\end{enumerate}
Since all inputs and outputs of the sub-systems are time continuous, it is reasonable to assume that also the coupling process is a time continuous element which includes the overall correlation between the continuous output signal of a sub-system and the continuous input signal of another sub-system. 
The coupling process includes two different effects.
On the one hand, the disturbing effects of sampling and delay due to data exchange and communication and, on the other hand, the added methods of compensation and reconstruction to compensate for these effects.  
Figure~\ref{fig:CouplingOverview} displays an overview of two distributively coupled systems ``A'' and ``B'' and the coupling element.

To enable a closed-loop analysis of the coupled system in frequency domain, the Laplace transform is applied to each component separately. 
The transfer function of the disturbing effects $G_f(s)$ describes the correlation between an output signal $y(t)$ and the ideally reconstructed input $u_s(t)$ of another sub-system in frequency domain.
Hence, under assumption 2) and 3) holds
\begin{equation}
U(s)=\underbrace{\frac{1}{\Delta T} \cdot e^{-s\tau}}_{G_f(s)} \cdot Y(s).
\label{Gf} 
\end{equation}
A detailed derivation on why the influence of the sampling can be represented by a scaling with $\frac{1}{\Delta T}$ under these circumstances can be found in~\cite{Benedikt.2013}. The authors carry out the Laplace transformation of a coupling process of a non-iterative offline co-simulation (without delay and compensation). 

The transfer function $G_{c}(s)$ represents the frequency domain correlation of the ideally reconstructed input $u_s(t)$ and the applied input $\hat{u}(t)$.
It holds
\begin{equation}
\hat{U}(s)=G_c(s) \cdot U(s).
\label{Gc} 
\end{equation}
$G_c(s)$ is now derived for a feedforward neural network with linear activation function as compensation and ZOH as the reconstruction method.
Starting point to get $G_c(s)$ is equation \eqref{ANN_func} which contains the sampled behavior of the compensation method under consideration of the constant time-delay $\tau$.
Including the ZOH reconstruction leads to a piece wise constant function in time domain
\begin{equation}
\hat{u}(t)=\vec{a}^T\vec{u} + b \text{\quad with \quad} n\Delta T \leq t < (n+1)\Delta T.
\label{timedomain} 
\end{equation}

The Laplace transform of $\hat{u}(t)$ is defined as
\begin{equation}
	\mathcal{L}\{\hat{u}(t)\}(s)= \sum_{n=0}^{\infty} \int_{n\Delta T}^{(n+1)\Delta T}\hat{u}(t)e^{-s\Delta T}dt.
	\label{laplace}
\end{equation}
The linearity property of the Laplace transform allows a separate transformation of each summand of $\hat{u}(t)$.
For the first summand $\hat{u}_1(t)=a_1 \cdot u_{t-\tau}$ equation \eqref{laplace} simplifies to
\begin{equation}
\mathcal{L}\{\hat{u}_1(t)\}(s)=\hat{U}_1(s)= \sum_{n=0}^{\infty} \int_{n\Delta T}^{(n+1)\Delta T}\hat{u}_1(t)e^{-s\Delta T} dt.
\label{laplace_simple} 
\end{equation}
Since $\hat{u}_1(t)$ is piece wise constant in, it holds
\begin{align}	
	\hat{U}_1(s)&= \sum_{n=0}^{\infty} \hat{u}_1(t) \frac{e^{-sn\Delta T}-e^{-s(n+1)\Delta T}}{s} \\
	&= \underbrace{a_1 \frac{1-e^{-s \Delta T}}{s}}_{G_{c1}(s)}\underbrace{\sum_{n=0}^{\infty}u_{t-\tau}e^{-sn \Delta T}}_{U_1(s)}.
	\label{}
\end{align}
This results in the Laplace transform of the first summand $G_{c1}(s)$ of $G_{c}(s)$, which is equal to a ZOH extrapolation scaled by $a_1$.
The transform for all other summands of $\hat{u}(t)$ works accordingly under consideration of an additional time shift. 
Therefore it holds
\begin{equation}
G_{c}(s) = \sum_{n=0}^{p-1} a_{n+1} e^{-ns\Delta T} \cdot \frac{1- e^{-s\Delta T}}{s}+\frac{b}{s}e^{-s\Delta T}
\end{equation}
for a compensation taking $p$ past signal values into account.
Finally, using equations \eqref{Gf} and \eqref{Gc}, the overall coupling process in frequency domain $G_e(s)$ can be written as
\begin{align}
&G_{p}(s)=G_{f}(s) \cdot G_{c}(s) \\
&= \sum_{n=0}^{p-1} a_{n+1} e^{-(\tau + n \Delta T)s} \cdot \frac{1- e^{-s\Delta T}}{s\Delta T}+\frac{b}{s\Delta T}e^{-(\tau + \Delta T)s}.
\label{eq_final} 
\end{align}
Remarks:
\begin{itemize}[noitemsep]
	\item The methodology also allows using other methods to reconstruct the input between macro time steps (e.g. FOH) by changing equation \eqref{timedomain} accordingly~\cite{Benedikt.2013}.
	\item When using a piece wise linear activation function in the neural network (e.g. ReLU or leaky ReLU) the behavior is nonlinear. Of course in this case the Laplace transform is not valid any more, but since the network will still maintain a piece wise linear structure, each linear component can be transformed separately, to examine the different possibilities of the transmission behavior of the neural network.
\end{itemize}
\subsection{Linear Two-Mass Oscillator}

Using the example of a two-mass oscillator, it is shown how the transfer function of the coupling process $G_{p}(s)$ can be used to estimate the overall system behavior of a distributively coupled system.
Here, this is done purely in simulation with synthetic coupling faults between the sub-systems.

\begin{figure}[h]
	\def\svgwidth{\columnwidth}
	\centerline{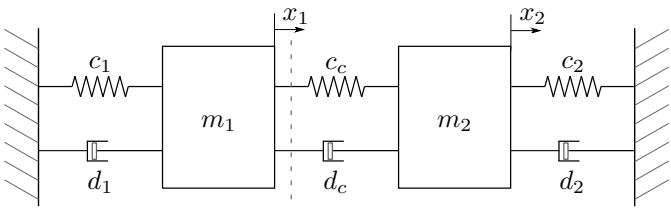}
	\caption{Linear Two-Mass Oscillator}
	\label{fig:Two Mass}
\end{figure}

Two via spring and damper connected models of single mass oscillators form the two-mass oscillator as can be seen in figure~\ref{fig:Two Mass}.
At the dashed line, the oscillator is divided into two sub-models each containing one mass. 
The models are coupled using the so called force/displacement coupling~\cite{Li.2017} approach, where one model calculates the coupling force and receives position and velocity of the mass of the other model.
Table~\ref{tab:TMO} contains the parameters of the slightly damped coupled system. 
\begin{table}[htbp]
	\renewcommand{\arraystretch}{1.3}
	\caption{Parameters of the Two-Mass Oscillator (SI)}
	\label{tab:TMO}
	\centering
	\begin{tabular}{c|c}
		\hline
		Parameter  & Value\\
		\hline
		\hline	
		$m_1, m2$   &   $100, 1$\\
		\hline
		$c_1$, $c_2$, $c_c$    &   $10$\\
		\hline
		$d_1$, $d_2$, $d_c$    &   $0.01$\\
		\hline
	\end{tabular}
\end{table}

\begin{figure}[htbp]
	\def\svgwidth{\columnwidth}
	\centerline{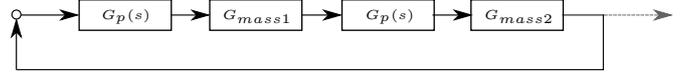}
	\caption{Interpretation of the System as Control Circuit.}
	\label{fig:ControlCircuit}
\end{figure}

A Laplace transform of the system equations of the two single mass oscillators~\cite{Li.2017} yields their transfer functions $G_{mass1}(s)$ and $G_{mass2}(s)$.
Together with the transfer function of the coupling process $G_{p}(s)$ the coupled system can be interpreted as a control circuit, which allows stability analysis with Bode and Nyquist plots. Figure~\ref{fig:ControlCircuit} sketches the idea. 

In order to determine $G_p(s)$ explicitly, numerical values must be assigned to the parameters of the distributed simulation and the compensation method. 
The macro step size is set to $\Delta T = 0.001s$ and the delay per coupling direction to $\tau = 0.003s$.
The resulting round-trip-time of $RTT=0.006s$ is much lower than e.g. in a coupling via the internet, but since the two-mass oscillator is a strongly coupled system, even this short round-trip-time has an effect on the system behavior.

The neural network which is used for compensating the delay is implemented in Python using the Keras package~\cite{chollet2015keras}. 
The size of the network and especially the number of inputs (considered past signal values) is a trade-off between computation resources needed and the capability of the network to represent nonlinear signal behavior. 
With a linear activation function of course, the capability of the network does not increase with the number of neurons~\cite{Dorffner.1996}, but for the time-delay compensation of a nonlinear signal a single neuron is not sufficient. 
The chosen network size of four input neurons, two hidden neurons and one output neuron is the smallest, which gives good results also for nonlinear signals (see nonlinear example).

For this example, the weights of the network are obtained by training the network before simulation on a self-created data set which is based on simulation results of the two-mass oscillator with different initial conditions (without faults).
Since this is cumbersome, section~\ref{design} presents how the information from the analysis shown here can be used to optimally parameterize the network in advance without training. 
After the training the parameters (equation \eqref{ANN_func}) calculated from the weights of the network read as 
\begin{equation}
\label{ANN_func_pre}
\vec{a} = 
\begin{bmatrix}
2.4748 \\
-0.6470 \\
-0.1664 \\
-0.6664 
\end{bmatrix},
b=0.
\end{equation}

Thus, the coupling process including compensation for this example can be determined by equation \eqref{eq_final}.
Multiplying the transfer functions, as shown in figure~\ref{fig:ControlCircuit}, results in the open-loop transfer function representation of the overall distributively coupled system
\begin{equation}
G_{sys}(s) = G_{mass2}(s) \cdot G_p(s) \cdot G_{mass1}(s) \cdot G_p(s).
\end{equation}

\begin{figure}[h!]
	\centering
	\resizebox{0.47\textwidth}{!}{\input{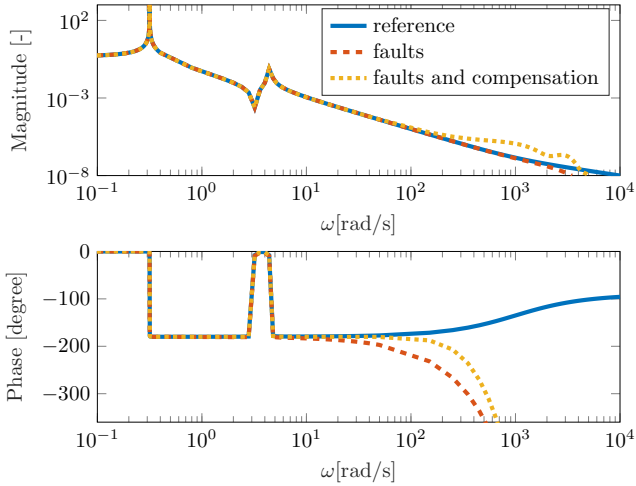}}
	\caption{Bode Plot of $G_{sys}(s)$}
	\label{fig:bode_pre}
\end{figure}

Figure~\ref{fig:bode_pre} shows the open loop Bode plot of $G_{sys}(s)$. The blue curve is the reference without coupling effects ($G_p(s)=1$), red curve is with included faults and trivial (ZOH) compensation and the yellow curve is with included faults and neural network compensation.
First of all, the two resonance frequencies of the two-mass oscillator are clearly visible and, more importantly, they do not change by adding faults or the compensation method. 
This means the qualitative system behavior stays the same.
The Bode plot is also useful to verify, that the frequency bandwidth of the system  (areas with high frequency amplification) is below $6\text{rad}/s$ and therefore small enough to avoid aliasing (assumption 2).
It is further observed, that the magnitude for high frequencies is larger when the compensation method is included in the control circuit. However, this amplification is not critical, since the two-mass oscillator itself dampens very strongly in this frequency range.
In the bode phase diagram a rapidly decreasing phase for high frequencies is visible, which is typical for systems with delays.
It is noticeable that without compensation the phase deviates away from the reference within the bandwidth, whereas with compensation the phase follows the reference in a larger frequency range.

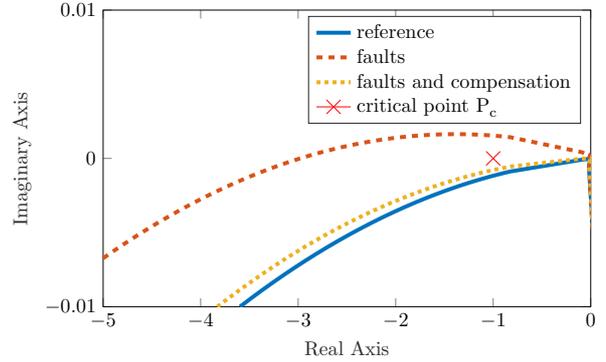
\begin{figure}[h!]
	\centering
	\setlength\figureheight{5.25cm} 
	\resizebox{0.47\textwidth}{!}{
%
%
\definecolor{mycolor1}{rgb}{0.00000,0.44700,0.74100}%
\definecolor{mycolor2}{rgb}{0.85000,0.32500,0.09800}%
\definecolor{mycolor3}{rgb}{0.92900,0.69400,0.12500}%
\begin{tikzpicture}

\begin{axis}[%
width=0.951\figurewidth,
height=\figureheight,
at={(0\figurewidth,0\figureheight)},
scale only axis,
unbounded coords=jump,
xmin=-5,
xmax=0,
xlabel style={font=\color{white!15!black}},
xlabel={Real Axis},
xtick = {-5,-4,-3,-2,-1,0},
ymin=-0.01,
ymax=0.01,
ylabel style={font=\color{white!15!black}},
ylabel={Imaginary Axis},
ytick = {-0.01,0,0.01},
tick label style={/pgf/number format/fixed, /pgf/number format/precision=5},
scaled ticks=false,
axis background/.style={fill=white},
legend style={legend cell align=left, align=left, draw=white!15!black}
]
\addplot [color=mycolor1, line width=2.0pt]
  table[row sep=crcr]{%
-3.59683215507419	-0.0100001840676875\\
-3.48486857220868	-0.00944266735261978\\
-3.37246772219028	-0.00889900166291469\\
-3.26032320516715	-0.00837258025631904\\
-3.14822719606642	-0.00786236204231905\\
-3.03610556341894	-0.00736800436418461\\
-2.92399862051459	-0.00688968691397029\\
-2.81171419033385	-0.00642662411619144\\
-2.69953769989311	-0.00598000744243032\\
-2.58724682278568	-0.00554895251968635\\
-2.47508929180059	-0.00513440559198886\\
-2.3629311163752	-0.00473584223600332\\
-2.25064977138008	-0.00435285259034535\\
-2.13837935450521	-0.0039859169553651\\
-2.02610229065883	-0.00363497563387138\\
-1.91392794229269	-0.00330034828641423\\
-1.80167944053049	-0.00298149920076396\\
-1.68937530640154	-0.00267850627930821\\
-1.57714493634512	-0.00239171165918384\\
-1.46491430125683	-0.00212090651133945\\
-1.35260870754716	-0.00186592157619181\\
-1.24025684598321	-0.00162684115571077\\
-1.1278781806266	-0.00140371316683963\\
-1.01549557108821	-0.00119657693511677\\
-0.903006850511927	-0.0010052496040589\\
-0.826612409282194	-0.000884430035002381\\
-0.0895316709210441	-8.74189765798228e-05\\
-0.00229501543928601	-2.56895813288693e-05\\
0.000242393714812383	-3.31399978481528e-05\\
nan	nan\\
0.0745243335214414	-0.0102682148723123\\
-0.0186796902900337	-0.000611200253904265\\
-0.00898250579283077	-0.000152171673563739\\
-0.00601430203003428	-7.73179542412628e-05\\
-0.00299315692750213	-3.02951819750596e-05\\
-6.28988454107748e-05	-2.53216103551779e-06\\
-4.01624830250746e-05	-2.01635488217278e-06\\
-9.99989868688544e-10	-1.00000021596713e-08\\
};
\addlegendentry{reference}

\addplot [color=mycolor2, dashed, line width=2.0pt]
  table[row sep=crcr]{%
-5.00024170368065	-0.00673416391073278\\
-4.88767323339142	-0.00622233258671034\\
-4.77537108884433	-0.00572777715116857\\
-4.66362569160005	-0.00525160110817247\\
-4.5518657813519	-0.00479125521555801\\
-4.43964920361227	-0.00434501980439972\\
-4.32740861507558	-0.00391471924052844\\
-4.21553633991759	-0.00350178508515953\\
-4.10370363497465	-0.00310491800131718\\
-3.99168410848445	-0.00272334746774128\\
-3.87934420582856	-0.00235672870245374\\
-3.76721033859925	-0.00200680589297875\\
-3.65520740485108	-0.00167327451137211\\
-3.54333013067414	-0.00135606616566797\\
-3.43115227399637	-0.00105401225690827\\
-3.31886916023282	-0.000767728496171571\\
-3.20671131274243	-0.000497801556983646\\
-3.09454068613463	-0.000243878718155344\\
-2.98234347302053	-5.94132414999393e-06\\
-2.87021085302222	0.000215820655511401\\
-2.75831976533748	0.000421117471335997\\
-2.64633046289994	0.00061059690441656\\
-2.53440464423653	0.000783970604175011\\
-2.42256344576811	0.000941229043989189\\
-2.31076504504569	0.00108244830771032\\
-2.19896785529092	0.00120767846137682\\
-2.08717934797462	0.00131689908248411\\
-1.9754905014037	0.00141002748108576\\
-1.8639463887686	0.00148705791036186\\
-1.75236812780014	0.00154811234900532\\
-1.6409199518436	0.00159309025174359\\
-1.52963594453282	0.00162200201895324\\
-1.41852684234293	0.00163486825788439\\
-1.30769095430232	0.00163171043421162\\
-1.19714841173208	0.00161256816304256\\
-1.08706027761063	0.00157752043185422\\
-0.977535154742072	0.00152667245573301\\
-0.868764477953896	0.00146019180768597\\
-0.826611538675433	0.00143008501381736\\
-0.0895307484725594	0.000413957091893735\\
-0.0344176070191162	0.000240985105921077\\
-0.017340979571606	0.00015987599230094\\
-0.00961483000266394	0.000106877029415031\\
-0.0052603342688915	6.34851043193052e-05\\
-0.00229507650806315	1.92947617234651e-05\\
0.000241590412188408	-3.85608140760496e-05\\
nan	nan\\
0.0700497765505306	-0.0119999999999996\\
-0.0186896468429589	1.66640896797077e-05\\
-0.00898207398866813	0.000174819747511989\\
-0.00601269637074253	0.00015844129749798\\
-0.0029911967209939	0.00011215917575047\\
-0.00102850330467419	6.2910477477196e-05\\
-0.000217399419890185	2.82242173383906e-05\\
-6.11409016988418e-05	1.49469110075628e-05\\
-3.841094103052e-05	1.18750443407478e-05\\
4.45095925449834e-07	7.23115575418376e-07\\
nan	nan\\
1.75202092833615e-08	-2.09549782859142e-07\\
3.56369689313851e-09	8.89980276141955e-08\\
nan	nan\\
2.25664216202404e-08	-5.0927257078115e-08\\
1.58185606835559e-08	2.66176547469854e-08\\
nan	nan\\
1.20312924067889e-08	-2.01111207687177e-08\\
6.0395342060815e-09	1.17607701355382e-08\\
nan	nan\\
5.34414734687516e-09	-8.12597011901062e-09\\
2.15930207048132e-09	4.38034142291599e-09\\
nan	nan\\
1.82000814419325e-09	-2.72391886824153e-09\\
5.49311707231936e-10	1.09743059084622e-09\\
nan	nan\\
3.52466500430637e-10	-5.40369526902396e-10\\
1.90854443360422e-10	3.17708526154092e-10\\
nan	nan\\
2.13099760060231e-10	-3.94881460863417e-10\\
2.89305468470502e-10	4.45356640454975e-10\\
nan	nan\\
2.35935715409141e-10	-4.77665906828406e-10\\
2.70053313045082e-10	3.83293397021589e-10\\
nan	nan\\
1.68374647557812e-10	-3.76622288911221e-10\\
-3.0794211625107e-10	-2.04479100318622e-10\\
};
\addlegendentry{faults}

\addplot [color=mycolor3, dotted, line width=2.0pt]
  table[row sep=crcr]{%
-3.82979405548606	-0.0100025210693566\\
-3.71824368549297	-0.00944437837675993\\
-3.6063997138168	-0.00890081109448593\\
-3.49484905659766	-0.008374671892454\\
-3.38316409178975	-0.00786390979457741\\
-3.27152839262389	-0.00736938278086452\\
-3.15974544553585	-0.00689024165870133\\
-3.04814067322872	-0.00642787491056795\\
-2.93668897884017	-0.00598210758868678\\
-2.82513434597201	-0.00555190531563987\\
-2.71340384832512	-0.00513704718265462\\
-2.60186373808038	-0.00473889042216769\\
-2.4903898182509	-0.00435693707926932\\
-2.37880813611407	-0.0039906007071564\\
-2.26718060962985	-0.00364011836144584\\
-2.15555800651834	-0.00330565913638825\\
-2.04402732089997	-0.00298746394441274\\
-1.93242158952824	-0.00268505412770192\\
-1.82079902195719	-0.00239860874556097\\
-1.70917142760638	-0.00212816476307465\\
-1.59762703627626	-0.00187392153699673\\
-1.48611433738459	-0.00163573997631117\\
-1.37459291548429	-0.0014135326766116\\
-1.26309871694398	-0.00120737229438372\\
-1.15159928649875	-0.00101720066090527\\
-1.04009431553952	-0.000843028437754523\\
-0.928632258474694	-0.000684935484003724\\
-0.81836926799956	-0.000544311969286504\\
-0.0886395507555928	-1.47789115505503e-05\\
-0.00952012770429622	-8.18838156524393e-06\\
-0.0022724746001539	-1.89946564961296e-05\\
0.000239908116648646	-3.35923821883632e-05\\
nan	nan\\
0.0737580162660518	-0.0104978115051013\\
-0.0185030027930737	-0.000515355594448597\\
-0.00889747868157409	-0.000103838408099577\\
-0.00595767840222283	-4.27724533582641e-05\\
-0.00296566018109434	-9.56276564378911e-06\\
-6.39376438873462e-05	2.03302675139128e-07\\
-4.14231272918286e-05	2.53436684527486e-07\\
4.4159428558288e-08	2.03171751245179e-06\\
nan	nan\\
2.40476452084692e-07	-1.32725630841435e-06\\
3.65037506000476e-07	5.44886388897936e-07\\
nan	nan\\
1.3637302087588e-07	-3.27219179752092e-07\\
1.09818588889254e-07	1.53412149206389e-07\\
nan	nan\\
7.79433610986757e-08	-1.97343307473119e-07\\
8.38656366575208e-08	1.51654858893124e-07\\
nan	nan\\
2.12787378828239e-08	-1.09923332214379e-07\\
1.16194804888892e-08	3.26495372959812e-08\\
nan	nan\\
7.25182491834175e-09	-1.50328887116302e-08\\
6.18787376893692e-09	8.55076365269269e-09\\
nan	nan\\
2.69011923847984e-10	-5.66770053112009e-09\\
1.32138744390886e-10	1.08445741275887e-09\\
nan	nan\\
2.67637911832708e-10	-1.8224088904617e-11\\
7.65646213096716e-10	5.15115505805852e-10\\
nan	nan\\
8.20512102706061e-10	-1.50681245258966e-09\\
1.34832855991363e-09	2.6800228702939e-09\\
nan	nan\\
1.25804877626479e-09	-2.40280373375867e-09\\
2.35763764067087e-09	3.00901525918107e-09\\
nan	nan\\
1.34476163538011e-09	-5.36040589693698e-09\\
3.96599642016326e-09	4.64812233147427e-09\\
nan	nan\\
3.66806141016696e-09	-3.64764485283331e-09\\
-3.26854943111243e-09	-1.37445610448594e-12\\
};
\addlegendentry{faults and compensation}

\addplot [color=red, draw=none, mark size=5.0pt, mark=x, mark options={solid, red}]
  table[row sep=crcr]{%
-1	0\\
};
\addlegendentry{$\text{critical point P}_\text{c}$}

\end{axis}

\begin{axis}[%
width=1.227\figurewidth,
height=1.227\figureheight,
at={(-0.16\figurewidth,-0.135\figureheight)},
scale only axis,
xmin=0,
xmax=1,
ymin=0,
ymax=1,
axis line style={draw=none},
ticks=none,
axis x line*=bottom,
axis y line*=left
]
\end{axis}
\end{tikzpicture}%
}
	\setlength\figureheight{7cm} 
	\caption{Enlargement of Nyquist Plot of $G_{sys}(s)$ with a pre-trained Neural Network as Compensation Method}
	\label{fig:nyquist_large_pre}
\end{figure}

How this affects stability can be investigated using the Nyquist plot in figure~\ref{fig:nyquist_large_pre}.
A close look at the area around the critical point $P_c=(-1, 0j)$ reveals a deviation of the Nyquist locus in the case without compensation.
The encirclement of the Nyquist locus around the critical point is different from the reference when the faults are introduced and changes back to the reference when the compensation is added (Nyquist Stability Criterion). 
This behavior is confirmed with simulation results: The system with faults is unstable without the compensation and can be stabilized by adding the compensation method. The reference system is of course stable, since a two-mass oscillator is a mechanically stable system.

This section revealed two important things:
Firstly, an analysis of the overall system behavior of a distributed real-virtual prototypes in the frequency domain is possible. 
This allows, for example, to check in advance whether a system can be stabilized under a certain time-delay with a certain compensation and reconstruction method.
Of course these statements are subject to some simplifications and assumptions and can therefore not be transferred one-to-one to reality, but the basic system behavior under certain coupling errors can be demonstrated with this analysis approach.
Secondly, it is shown that the trained neural network is able to compensate for the coupling faults.

\section{DESIGN OF THE COMPENSATION METHOD}
\label{design}
The idea is to use the analysis method from the last section not only for checking the system behavior under the influence of delays and a compensation method, but also for an optimal design of the compensation method itself.
This eliminates the need for training the neural network in beforehand.

The compensation method is very flexible due to the parameters $\vec{a}$ and $b$. 
Therefore, requirements for the behavior of the compensation in the frequency domain are made first, before on that basis an optimization problem is defined, by whose solution the optimal parameters are found.
Since the compensation depends strongly on the configuration and properties of the coupled systems, the requirements are defined for the transfer function of the overall coupling process $G_p(s,\vec{a},b)$.

The parameters $\vec{a}$ and $b$ must be chosen such that
\begin{itemize}[noitemsep]
	\item $\sum_{p}a_p+b=1$, which leads to $\lim_{w \to 0} G_p(s=j\omega)= 1$ and thus a correct extrapolation of constant signals.
	\item the magnitude of $G_p(s)$ within the bandwidth of the coupled system is neutral ($|G_p(j\omega)| = 1$).
	\item the phase shift of $G_p(s)$ within the bandwidth of the coupled system is neutral ($\angle G_p(j\omega)=0^\circ$).
	\item outside the bandwidth frequencies of the coupled system, the combined magnitude of all coupling processes $G_p(s)$ does not increase faster than the magnitude of the coupled system decreases. This guarantees that additional amplifications of $G_{sys}(s)$, which are introduced by $G_p(s)$, will be damped by the dynamics of the coupled system itself.
\end{itemize}
The second and third requirement ensure the compensation of the coupling faults in frequency ranges where the coupled system is dynamically active.

All requirements are weighted and combined into a single objective function, but a multi-criteria optimization would also be possible.
The combined objective function of the optimization problem with $s=j\omega$ reads as
\begin{align}
J(a,b) &= \alpha J_a + \beta J_p + \gamma J_r \text{, with}\\
J_a &= \int_{\omega_{bw, min}}^{\omega_{bw,max}} \frac{1-|G_p(j\omega)|}{\omega_{bw,max}-\omega_{bw, min}} d\omega \\
J_p &= \int_{\omega_{bw, min}}^{\omega_{bw,max}} \frac{\angle G_p(j\omega)}{\omega_{bw,max}-\omega_{bw, min}} d\omega \\
J_r &= \int_{0}^{\omega_{bw,min}}  \max \left(|G_p(j\omega),1  \right) d\omega - \omega_{bw, min}\\
&+ \int_{\omega_{bw,max}}^{\frac{2 \pi}{\Delta T}} \max \left(|G_p(j\omega)|- \left(\frac{\omega}{\omega_{bw,max}}\right)^v,0 \right) d\omega.
\label{objective function} 
\end{align}
The bandwidth frequencies of the coupled system are within the interval $[\omega_{bw, min}, \omega_{bw,max}]$ and the remaining frequencies (up to the sampling frequency) in $[0, \omega_{bw,min})$ and $(\omega_{bw, max}, \frac{2 \pi}{\Delta T} ]$.
The optimization problem results in
\begin{equation}
\label{opt_prob}
	\min_{a,b} J(a,b) \text{ such that } \sum_{p}a_p+b=1.
\end{equation}

Remarks:
\begin{enumerate}[noitemsep]
	\item It shall apply $\gamma >> \alpha, \beta$, to ensure that the coupling process stays inside the magnitude boundary for frequencies outside the bandwidth frequencies. Furthermore holds $\alpha=100 \beta$ to punish a phase difference of $1^\circ$ equally as a magnitude error of $1\%$ \cite{Benedikt.2013}.
	\item Parameter $v$ depends on the relative degree $r$ of the coupled system. It holds $v=\frac{1}{2r}$ 
	\item The calculation of the objective function is possible with very little system information. Transfer functions of the coupled subsystems are not necessary. The bandwidth of the coupling signals could instead be estimated by Fourier transformations of the coupling signals and the relative degree can be set conservatively to one in case of doubt.	
\end{enumerate}

\begin{figure}[h!]
	\centering
	\resizebox{0.47\textwidth}{!}{\input{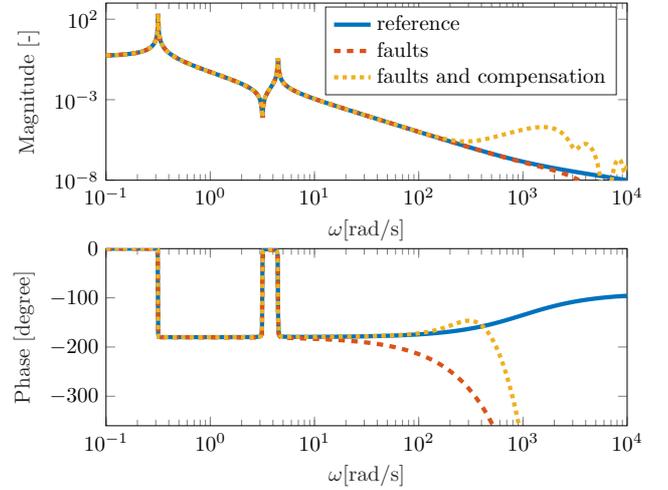}}
	\caption{Bode Plot of $G_{sys}$ with an Optimally Initialized Neural Network as Compensation}
	\label{fig:bode_opt}
\end{figure}

\begin{figure}[h!]
	\centering
	\setlength\figureheight{5.25cm} 
	\resizebox{0.47\textwidth}{!}{
%
%
\definecolor{mycolor1}{rgb}{0.00000,0.44700,0.74100}%
\definecolor{mycolor2}{rgb}{0.85000,0.32500,0.09800}%
\definecolor{mycolor3}{rgb}{0.92900,0.69400,0.12500}%
\begin{tikzpicture}

\begin{axis}[%
width=0.951\figurewidth,
height=\figureheight,
at={(0\figurewidth,0\figureheight)},
scale only axis,
unbounded coords=jump,
xmin=-5,
xmax=0,
xlabel style={font=\color{white!15!black}},
xlabel={Real Axis},
xtick = {-5,-4,-3,-2,-1,0},
ymin=-0.01,
ymax=0.01,
ylabel style={font=\color{white!15!black}},
ylabel={Imaginary Axis},
ytick = {-0.01,0,0.01},
tick label style={/pgf/number format/fixed, /pgf/number format/precision=5},
scaled ticks=false,
axis background/.style={fill=white},
legend style={legend cell align=left, align=left, draw=white!15!black}
]
\addplot [color=mycolor1, line width=2.0pt]
  table[row sep=crcr]{%
-3.59683215507419	-0.0100001840676875\\
-3.48486857220868	-0.00944266735261978\\
-3.37246772219028	-0.00889900166291469\\
-3.26032320516715	-0.00837258025631904\\
-3.14822719606642	-0.00786236204231905\\
-3.03610556341894	-0.00736800436418461\\
-2.92399862051459	-0.00688968691397029\\
-2.81171419033385	-0.00642662411619144\\
-2.69953769989311	-0.00598000744243032\\
-2.58724682278568	-0.00554895251968635\\
-2.47508929180059	-0.00513440559198886\\
-2.3629311163752	-0.00473584223600332\\
-2.25064977138008	-0.00435285259034535\\
-2.13837935450521	-0.0039859169553651\\
-2.02610229065883	-0.00363497563387138\\
-1.91392794229269	-0.00330034828641423\\
-1.80167944053049	-0.00298149920076396\\
-1.68937530640154	-0.00267850627930821\\
-1.57714493634512	-0.00239171165918384\\
-1.46491430125683	-0.00212090651133945\\
-1.35260870754716	-0.00186592157619181\\
-1.24025684598321	-0.00162684115571077\\
-1.1278781806266	-0.00140371316683963\\
-1.01549557108821	-0.00119657693511677\\
-0.903006850511927	-0.0010052496040589\\
-0.826612409282194	-0.000884430035002381\\
-0.0895316709210441	-8.74189765798228e-05\\
-0.00229501543928601	-2.56895813288693e-05\\
0.000242393714812383	-3.31399978481528e-05\\
nan	nan\\
0.0745243335214414	-0.0102682148723123\\
-0.0186796902900337	-0.000611200253904265\\
-0.00898250579283077	-0.000152171673563739\\
-0.00601430203003428	-7.73179542412628e-05\\
-0.00299315692750213	-3.02951819750596e-05\\
-6.28988454107748e-05	-2.53216103551779e-06\\
-4.01624830250746e-05	-2.01635488217278e-06\\
-9.99989868688544e-10	-1.00000021596713e-08\\
};
\addlegendentry{reference}

\addplot [color=mycolor2, dashed, line width=2.0pt]
  table[row sep=crcr]{%
-5.00024170368065	-0.00673416391073278\\
-4.88767323339142	-0.00622233258671034\\
-4.77537108884433	-0.00572777715116857\\
-4.66362569160005	-0.00525160110817247\\
-4.5518657813519	-0.00479125521555801\\
-4.43964920361227	-0.00434501980439972\\
-4.32740861507558	-0.00391471924052844\\
-4.21553633991759	-0.00350178508515953\\
-4.10370363497465	-0.00310491800131718\\
-3.99168410848445	-0.00272334746774128\\
-3.87934420582856	-0.00235672870245374\\
-3.76721033859925	-0.00200680589297875\\
-3.65520740485108	-0.00167327451137211\\
-3.54333013067414	-0.00135606616566797\\
-3.43115227399637	-0.00105401225690827\\
-3.31886916023282	-0.000767728496171571\\
-3.20671131274243	-0.000497801556983646\\
-3.09454068613463	-0.000243878718155344\\
-2.98234347302053	-5.94132414999393e-06\\
-2.87021085302222	0.000215820655511401\\
-2.75831976533748	0.000421117471335997\\
-2.64633046289994	0.00061059690441656\\
-2.53440464423653	0.000783970604175011\\
-2.42256344576811	0.000941229043989189\\
-2.31076504504569	0.00108244830771032\\
-2.19896785529092	0.00120767846137682\\
-2.08717934797462	0.00131689908248411\\
-1.9754905014037	0.00141002748108576\\
-1.8639463887686	0.00148705791036186\\
-1.75236812780014	0.00154811234900532\\
-1.6409199518436	0.00159309025174359\\
-1.52963594453282	0.00162200201895324\\
-1.41852684234293	0.00163486825788439\\
-1.30769095430232	0.00163171043421162\\
-1.19714841173208	0.00161256816304256\\
-1.08706027761063	0.00157752043185422\\
-0.977535154742072	0.00152667245573301\\
-0.868764477953896	0.00146019180768597\\
-0.826611538675433	0.00143008501381736\\
-0.0895307484725594	0.000413957091893735\\
-0.0344176070191162	0.000240985105921077\\
-0.017340979571606	0.00015987599230094\\
-0.00961483000266394	0.000106877029415031\\
-0.0052603342688915	6.34851043193052e-05\\
-0.00229507650806315	1.92947617234651e-05\\
0.000241590412188408	-3.85608140760496e-05\\
nan	nan\\
0.0700497765505306	-0.0119999999999996\\
-0.0186896468429589	1.66640896797077e-05\\
-0.00898207398866813	0.000174819747511989\\
-0.00601269637074253	0.00015844129749798\\
-0.0029911967209939	0.00011215917575047\\
-0.00102850330467419	6.2910477477196e-05\\
-0.000217399419890185	2.82242173383906e-05\\
-6.11409016988418e-05	1.49469110075628e-05\\
-3.841094103052e-05	1.18750443407478e-05\\
4.45095925449834e-07	7.23115575418376e-07\\
nan	nan\\
1.75202092833615e-08	-2.09549782859142e-07\\
3.56369689313851e-09	8.89980276141955e-08\\
nan	nan\\
2.25664216202404e-08	-5.0927257078115e-08\\
1.58185606835559e-08	2.66176547469854e-08\\
nan	nan\\
1.20312924067889e-08	-2.01111207687177e-08\\
6.0395342060815e-09	1.17607701355382e-08\\
nan	nan\\
5.34414734687516e-09	-8.12597011901062e-09\\
2.15930207048132e-09	4.38034142291599e-09\\
nan	nan\\
1.82000814419325e-09	-2.72391886824153e-09\\
5.49311707231936e-10	1.09743059084622e-09\\
nan	nan\\
3.52466500430637e-10	-5.40369526902396e-10\\
1.90854443360422e-10	3.17708526154092e-10\\
nan	nan\\
2.13099760060231e-10	-3.94881460863417e-10\\
2.89305468470502e-10	4.45356640454975e-10\\
nan	nan\\
2.35935715409141e-10	-4.77665906828406e-10\\
2.70053313045082e-10	3.83293397021589e-10\\
nan	nan\\
1.68374647557812e-10	-3.76622288911221e-10\\
-3.0794211625107e-10	-2.04479100318622e-10\\
};
\addlegendentry{faults}

\addplot [color=mycolor3, dotted, line width=2.0pt]
  table[row sep=crcr]{%
-3.59735777818835	-0.0100016312831968\\
-3.48536295819737	-0.00944392178584952\\
-3.37293170869504	-0.00890007603772158\\
-3.2607578197367	-0.00837348810081906\\
-3.14863340791679	-0.00786311608450463\\
-3.03648432269599	-0.00736861683152767\\
-2.92435088738259	-0.00689016962690969\\
-2.81204088163878	-0.00642698835237576\\
-2.69983979839865	-0.00598026441975819\\
-2.58752526261449	-0.00554911275820658\\
-2.47534505904985	-0.00513447938486911\\
-2.36293037079553	-0.00473502184868524\\
-2.25064908749953	-0.00435206804774824\\
-2.13837872962654	-0.00398516822729222\\
-2.02610172209412	-0.00363426272206269\\
-1.91392742730269	-0.00329967113799867\\
-1.80167897646158	-0.002980857890587\\
-1.68937489058692	-0.00267790082897612\\
-1.57736286254181	-0.0023916842846945\\
-1.46510406972079	-0.00212081784437679\\
-1.35285413921344	-0.00186596406750317\\
-1.24053548388296	-0.00162695283317671\\
-1.12817045890449	-0.00140384722410847\\
-1.01583308358232	-0.00119678562174697\\
-0.903474061655024	-0.00100565791556795\\
-0.826612274379899	-0.000884101408761495\\
-0.0895316504787211	-8.73492471864523e-05\\
-0.00229501002137589	-2.56851990978468e-05\\
0.00024239290175343	-3.31403387656692e-05\\
nan	nan\\
0.0745238372318195	-0.0102682019122851\\
-0.0186795664400297	-0.000611192933917426\\
-0.00898243471549076	-0.000152176527532966\\
-0.00601424666462425	-7.73276820718038e-05\\
-0.00299311638746413	-3.03137246184804e-05\\
-6.28875627946002e-05	-2.69548951736809e-06\\
-4.01616386240811e-05	-2.22042058517502e-06\\
-2.2454135177874e-06	-1.32907349170708e-06\\
1.83364821548793e-06	8.51377499122563e-06\\
nan	nan\\
5.38958613605089e-06	-1.37230138799183e-05\\
-8.34933271631755e-06	-1.42511026997028e-05\\
-1.78745875474284e-05	-2.14392295383448e-06\\
-1.30919644734107e-05	1.39305090796071e-05\\
3.81069801624889e-06	1.94259116779172e-05\\
nan	nan\\
7.89582036642145e-07	-1.89698065327093e-05\\
-1.44630579126392e-05	-1.04271118779309e-05\\
-1.52052830895144e-05	5.94568773149007e-06\\
-2.75392135895558e-06	1.43001071521986e-05\\
9.31903472478623e-06	8.53495937169768e-06\\
nan	nan\\
1.96895040094347e-06	-8.47059744613077e-06\\
-4.86066949800801e-06	1.91595974374792e-06\\
8.3524874572305e-08	1.37480153794556e-06\\
nan	nan\\
1.19693487565797e-06	-1.21881751624642e-06\\
5.43510513484335e-07	1.54656486417437e-06\\
nan	nan\\
8.92198392055121e-07	-9.4281071261193e-07\\
5.33577053651868e-08	5.34912622640604e-07\\
nan	nan\\
2.23631730911222e-08	-1.82093852441056e-07\\
8.44843617286983e-09	1.79313217607557e-08\\
nan	nan\\
4.54476012379246e-10	-1.26025456737011e-09\\
1.06186890391768e-08	1.16019420737246e-08\\
nan	nan\\
2.00683429874005e-08	-3.34279031122264e-08\\
2.3803978788095e-08	1.18717573993621e-07\\
nan	nan\\
1.34208832047733e-07	-9.53232706102369e-08\\
2.30595631478536e-08	8.07609619180027e-08\\
nan	nan\\
1.23010428509929e-08	-4.50480204250425e-08\\
3.13708263810497e-08	6.20696973818724e-08\\
};
\addlegendentry{faults and compensation}

\addplot [color=red, draw=none, mark size=5.0pt, mark=x, mark options={solid, red}]
  table[row sep=crcr]{%
-1	0\\
};
\addlegendentry{$\text{critical point P}_\text{c}$}

\end{axis}
\end{tikzpicture}%
}
	\setlength\figureheight{7cm} 
	\caption{Enlargement of Nyquist Plot of $G_{sys}$ with an Optimally Initialized Neural Network as Compensation}
	\label{fig:nyquist_large_opt}
\end{figure}
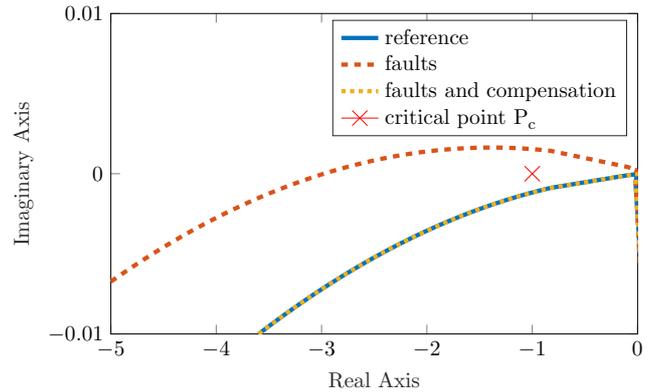

The bandwidth of the two mass oscillator is in the range $[1 \frac{rad}{s}, 6 \frac{rad}{s}]$ and the relative degree is $r=2$.
Thus, the numerically found solution of equation \eqref{opt_prob} is
\begin{equation}
\label{a_opt}
\vec{a}_{opt} = 
\begin{bmatrix}
6.5103 \\
-1.5509 \\
-9.9296 \\
5.9702 
\end{bmatrix},
b_{opt}=0.
\end{equation}
Figure \ref{fig:bode_opt} and \ref{fig:nyquist_large_opt} show the bode plot and the enlarged nyquist plot with the neural network compensation, initialized with the optimized $\vec{a}_{opt}$ and $b_{opt}$. Similar results as in the trained version of the neural network are obtained. The deviation of the nyquist curve between the reference and the compensated version is even smaller than with the trained neural network.

\section{NONLINEAR EXAMPLE}
\label{nonlinear-example}
In the last section, an optimal time-delay compensation of the form of equation \eqref{ANN_func} was designed on the basis of very little system information. 
Now the question arises why neural networks should be used as compensation method at all, since they are more complex to calculate and implement than equation \eqref{ANN_func} (and with linear activation function they behave exactly alike). 
The advantages of the neural network is that it can be adapted online by training in parallel and is also capable of represent nonlinear signal behavior. 

To show this, a nonlinearity extends the two-mass oscillator system from figure \ref{fig:Two Mass}. 
A mechanical stop prevents the first mass from positioning beyond $x_{1,stop}=-0.1m$. 
This is implemented by reversing the velocity of the first mass $v' = -e \cdot v$ at $x_{1,stop}$~\cite{Gall.2001}.
The coefficient of restitution is $e=0.7$, which leads to a partly inelastic collision.
For $x_1 > x_{1,stop}$  the system behaves linear and corresponds exactly to the two-mass oscillator from section~\ref{analysis} (all system parameters are identical), whereby the optimized compensation parameters from equation \eqref{a_opt} are also optimal in the nonlinear system.

\begin{figure}[h!]
	\centering
	\setlength\figureheight{4.9cm} 
	\resizebox{0.47\textwidth}{!}{\input{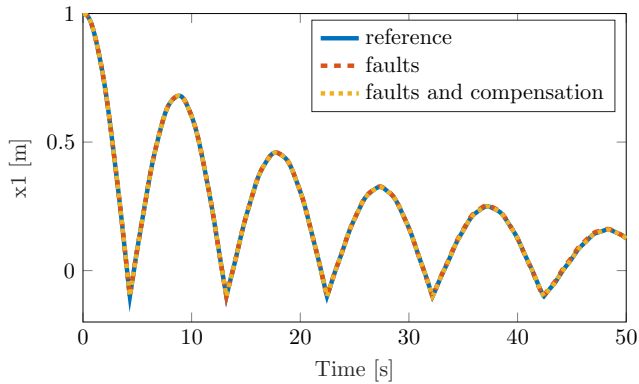}}
	\setlength\figureheight{7cm} 
	\caption{Position of the First Mass for $t\in[0s, 50s]$}
	\label{fig:x1_beginning}
\end{figure}

Figure~\ref{fig:x1_beginning} shows simulation results with the nonlinear two-mass oscillator.
The oscillation starts from the initial condition $x_{1,0} = x_{2,0} = 1$ and no external forces act on the system.  
The bouncing of the first mass is visible and all curves lie on top of each other, which means that the coupling faults have no significant effect on $x_1$ during the first $50s$ of the simulation.

\begin{figure}[h!]
	\centering
	\setlength\figureheight{4.9cm} 
	\resizebox{0.47\textwidth}{!}{\input{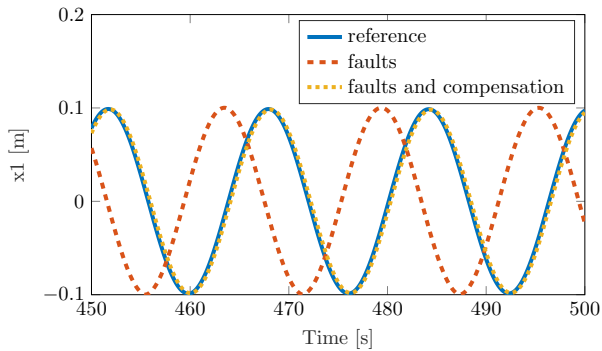}}
	\setlength\figureheight{7cm} 
	\caption{Position of the First Mass for $t\in[450s, 500s]$}
	\label{fig:x1_end}
\end{figure}

Figure~\ref{fig:x1_end}, which shows the same simulations $400s$ later, reveals a large deviation of the simulation results and also different convergence properties of the coupled system.
The slight damping reduces the oscillation amplitude of $x_1$ in each period in the reference simulation as well as in the simulation with faults and compensation. 
In the end all the energy will be dissipated and the system will end up in its equilibrium.
In the simulation with faults and without compensation, the oscillation amplitude of $x_1$ stays at $0.1m$ and the mass hits the mechanical stop in each period.
The energy fed into the system each period due to the instability caused by the faults (figure ~\ref{fig:nyquist_large_pre}) is dissipated by the partly elastic collision with the mechanical stop.

\begin{figure}[h!]
	\centering
	\setlength\figureheight{4.9cm} 
	\resizebox{0.47\textwidth}{!}{\input{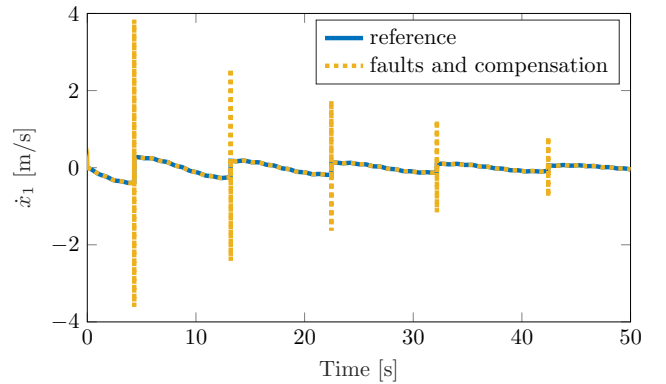}}
	\setlength\figureheight{7cm} 
	\caption{Velocity of the First Mass with Linear Neural Network as Compensation}
	\label{fig:x1_dot_comp}
\end{figure}

Figure~\ref{fig:x1_dot_comp} shows, however, the problem of the compensation method designed on the linear system. 
The velocity reversal at the mechanical stop causes jumps in the coupling signal $\dot{x}_1$, which leads to an overshoot by factor $a_{1,opt}=6.5103$ of the compensation.
In this particular case, the large compensation errors at the jumps do not effect the overall system behavior significantly (system is still stable, see figure~\ref{fig:x1_end}), but in general such large errors should be avoided.

Here the advantages of the neural network can be fully utilized. 
Instead of a linear activation function, leaky ReLUs are now used in the two neurons of the hidden layer, which enables the network to switch between four different behaviors.
The network is initialized the same way as the linear neural network from the previous example, but is now adapted during the simulation.
The online training is performed in parallel in an external process, in order not to influence the computation time of the compensation.
For this purpose, the values of the sampled coupling signal are stored during the simulation and sent to the training process together with the current configuration of the neural network. 
There, training data is created from the data points.
Each training sample consist of an input vector $\vec{x}$ with $p$ consecutive data points and a response $y$, which contains the data point $k=\frac{\tau}{\Delta T}$ steps after the latest value of the input vector.
The neural network from the compensation process is then duplicated and optimized in the training process, to minimize the cost 
\begin{equation}
	C(w)=\frac{1}{N} \sum_{i=1}^{N}L(f(x_i,y_i,w))
\end{equation}
for $N$ training samples.
The function $f$ stands for the neural network with the weights $w$. The loss function $L$ implements the mean-squared-error algorithm and the optimizer Adam~\cite{Kingma.2014}, a gradient-descent algorithm with adaptive learning rate, solves the optimization problem.
Once the optimization is complete, the updated weights are sent to the compensation process, where the neural network is updated.
This online adaption process can be repeated several times during a simulation.

\begin{figure}[h!]
	\centering
	\setlength\figureheight{4.9cm} 
	\resizebox{0.47\textwidth}{!}{\input{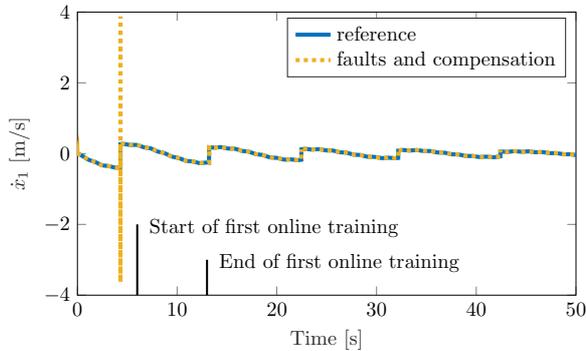}}
	\setlength\figureheight{7cm} 
	\caption{Velocity of the First Mass with Nonlinear and Online Adapting Neural Network as Compensation}
	\label{fig:x1_dot_comp_learn}
\end{figure}

Figure~\ref{fig:x1_dot_comp_learn} shows the simulation results of $\dot{x}_1$ again, but this time the online training was active. 
After the first jump at $4.5s$, the first online training cycle starts and finishes a few seconds later. 
This way, the network is able to switch its behavior at the second jump and the predicted signal does not overshoot anymore.
If the simulation would now be carried out using the adapted neural network as compensation method right from the start, also no overshoot would occur at the first jump.

\section{CONCLUSION AND FUTURE WORK}
\label{conclusion}
Current trends in the automotive sector require distributed mixed real-virtual test approaches.
In this work a feedforward neural network is used as a generalized compensation method for the coupling faults (constant time-delay) of mixed real-virtual prototypes. 
In order to be able to make statements in advance on how the compensation method will behave in a spatially distributed co-simulation, an analysis method in frequency domain is proposed that describes the overall coupling process.
Even with very little information on the coupled subsystems, it is possible to use the analysis to determine the parameters of the neural network via optimization in such a way that the coupling faults are compensated optimally.
Furthermore, it is shown that the neural network based compensation is also able to adapt to nonlinear signal behavior using an online learning strategy.

Future work will on the one hand focus on the consideration of different fault effects to be able to analyze the coupling process in a more realistic way. On the other hand the entire methodology will be tested on a real, industrial test bench coupled to a MiL simulation.

\bibliographystyle{plain}
\small {\bibliography{eurosis}}

\end{document}